# Machine Learning Algorithms for Active Monitoring of High Performance Computing as a Service (HPCaaS) Cloud Environments

Gianluca Longoni, Ryan LaMothe, Jeremy Teuton
Pacific Northwest National Laboratory
902 Battelle Boulevard
Richland, WA 99352

Mark Greaves, Nicole Nichols, William Smith
Pacific Northwest National Laboratory
1100 Dexter Ave N, Suite 400
Seattle, WA 98109

*Abstract*— Cloud computing provides ubiquitous and on-demand access to vast reconfigurable resources that can meet any computational need. Many service models are available, but the Infrastructure as a Service (IaaS) model is particularly suited to operate as a high performance computing (HPC) platform, by networking large numbers of cloud computing nodes. We used the Pacific Northwest National Laboratory (PNNL) cloud-computing environment to perform our experiments. A number of cloud computing providers such as Amazon Web Services, Microsoft Azure, or IBM Cloud, offer flexible and scalable computing resources. This paper explores the viability identifying types of engineering applications running on a cloud infrastructure configured as an HPC platform using privacy preserving features as input to statistical models. The engineering applications considered in this work include MCNP6, a radiation transport code developed by Los Alamos National Laboratory, OpenFOAM, an open source computational fluid dynamics code, and CADO-NFS, a numerical implementation of the general number field sieve algorithm used for prime number factorization. Our experiments use the OpenStack cloud management tool to create a cloud HPC environment and the privacy preserving Ceilometer billing meters as classification features to demonstrate identification of these applications.

## I. Introduction

The development of commercial cloud-based infrastructures such as Amazon Web Services, Microsoft Azure, and others, has transformed the computing business, moving it from the era of isolated mainframes in the '70s and '80s, through Beowulf cluster computing in the '90s and early 2000s, to the present-day cloud computing model, in which computational resources are offered as an on-demand service. The cloud model is a cost-effective solution for computing users—no significant capital investments are necessary to obtain computing resources and users can schedule resources based on short-term forecasts of their computing needs (elastic infrastructure). Several categories of users co-exist in the cloud space, including web hosting, social networks, and API providers.

However, due to the economical and abundant availability of computing power, a different category of cloud users is emerging, tied to the heavy-duty needs of scientific computing. In response, a new category of cloud-based service is maturing in the form of High Performance Computing as a Service (HPCaaS). HPCaaS includes the use of clustered servers and storage as resource pools, a web interface for users to submit job requests, and a smart scheduling mechanism that can schedule multiple applications simultaneously on a given cluster, taking into consideration the different application characteristics to achieve maximum efficiency.

While this new capacity presents emerging possibilities for a suite of research that has been difficult and/or expensive to conduct in the past, it may also bring a need for new metrics.This paper describes novel work that supports the identification of engineering applications running on an HPCaaS type of cloud infrastructure.

Similar work is discusseed by Nichols et al.:[1] the concept of which is to utilize telemetry data from a standard cloud-data collection service. The telemetry data is then used to train a set of machine learning (ML) classifiers in supervised mode to detect/recognize the applications running in its environment. In our work, we selected OpenStack as the cloud infrastructure management tool and Ceilometer (an OpenStack service) as the data collection infrastructure used to obtain the telemetry data. Ceilometer is capable of recording several activity indicators, such as CPU and memory utilization, disk I/O and network activity.[1] We use a Python script to create a pipeline from Ceilometer to RabbitMQ and to obtain a time-dependent snapshot of cloud activity. The ML infrastructure is based on the *scikit-learn* Python package, allowing flexible scripting capabilities.[2] The engineering applications considered in this work include the Monte Carlo N-Particle radiation transport code (MCNP6)[3]; the Open Field Operation And Manipulation (OpenFOAM) platform;[4] and CADO-NFS, an implementation of the general number field sieve algorithm.[5]

### A. Description of the Applications

MCNP6 is a Monte Carlo radiation transport computer code used for nuclear design, and it solves for the spatial and energy distributions of neutral and charged particles, e.g., neutrons, photons, and electrons. MCNP6 has many uses, including, assessing the dose absorbed from a radioactive source.

OpenFOAM is a platform used to solve systems of partial differential equations (PDEs) arising from engineering problems such as fluid flow, thermal condution, and structural behavior. Both MCNP and OpenFOAM are fully parallelized using the Message Passing Interface (MPI), or alternatively OpenMP for shared memory systems. Since MCNP is based on a Monte Carlo or stochastic approach, the parallelization is straightforward, involving partitioning the number of histories or random walks among the available processors. Because the random walks are inherently independent, and since the particles do not interact among each other but only with the surrounding environment, they can be treated independently by each processor. A synchronization step is only required to compute global statistical parameters such as means, standard deviations, and so on. This step is in general achieved with synchronous communication among processors using MPI's *Allreduce/Allgather* instructions. Due to the small amount of intra-processor communication involved, it is often observed that Monte Carlo computations are inherently parallel, which will be evident in the telemetry data acquired.

OpenFOAM's parallelization strategy is based on a domain partitioning algorithm. The discretized spatial domain is partitioned and allocated on each processor independently. The synchronization step in this case is more involved, since the processors need to communicate the solution on the sub-domain boundaries at each iteration. As with MCNP6, this aspect of OpenFOAM's parallelization will be evident in the telemetry data on network activity. Both MCNP and OpenFOAM operate using the main objectives of a parallel algorithm, namely parallel tasking, parallel memory, and parallel I/O. The parallelization of CADO-NFS is also based on the MPI model, but in this paper we have used a multithreaded version of the code.

The very complex nature of the problems solved, along with the very different computational architecture of each application we consider, provide a very diverse dataset to be used for the design of realistic cloud-monitoring algorithms. Also, the telemetry data obtained from the cloud is completely anonymous and not itself traceable to a particular user. This aspect is very important in the development of privacy-preserving monitoring tools.

Our paper is organized into five sections. Section II describes our cloud infrastructure and methods for acquiring telemetry data, including pipelines and pre-processing scripts. Section III discusses the telemetry data acquired for CADO-NFS, MCNP6, and OpenFOAM. Section IV presents the ML algorithms implemented in this study and the results obtained from our application detection experiments. This section will also discuss the application of Principal Component Analysis (PCA) to the data sets and its use to perform regularization of the data. Section V contains final remarks on the current results and future work to be performed.

This paper illustrates the building blocks required to create a signature database for a large number of applications of interest and describes the ML algorithms needed to perform real-time detection of applications running in a cloud environment configured as an HPCaaS system. Our future goal is to create an autonomous detection system capable of providing a privacy-preserving probabilistic metric for the detection of user/application activities in a cloud environment.

II. CLOUD INFRASTRUCTURE AND DATA ACQUISITION

This section describes the cloud infrastructure and tools we used to monitor our cloud environment and to obtain the telemetry data for the various applications. For this work, we obtained a number of virtual machines (VMs) and configured them as independent nodes, resulting in aconfiguration very similar to what can be achieved with off-the-shelf hardware used to build a Beowulf-type cluster. Specifically, two nodes with eight computing cores each are configured as a distributed memory environment. The physical compute nodes are dual-socket AMD Opteron 6272, with each CPU having 16 physical/32 logical cores, a clock frequency of 2.1 GHz, 512 KB of L3 cache memory, and a total of 128 GB RAM (64 GB RAM per socket). Local storage consists of 2 TB mirrored via software RAID 1. The cloud VMs are interconnected with a 40-Gb/sec network backbone.

We configured the two VM nodes, called *digi-a* and *digi-b*, to work as a cluster, with *digi-a* as the master node and *digi-b* as the slave node. The network file system (NFS) was reconfigured to share a drive partition between the two nodes, and the */etc/hosts* file was modified to be able to see the two nodes concurrently on the local area network (LAN).

This type of configuration is typical of high-performance computing (HPC) infrastructures and can be extended to several computing nodes/VMs. We found that once the VMs have been attained by the cloud management system, it is straightforward to configure the cloud environment as an HPC cluster. The MPI and OpenMP parallel libraries work seamlessly in this kind of configuration. The MPI library was compiled from scratch on the target architecture to take advantage of the "hardware sensing" features during the configuration process and to obtain the most optimized version of the library for the hardware at hand.

The OpenFOAM, MNCP, and CADO-NFS computer codes were also compiled from scratch using the MPI library wrappers coupled with the C++ and Fortran GNU compilers (v5.3.0).

OpenStack Ceilometer was used to collect the data produced by OpenStack services; The OpenStack telemetry features that Ceilometer can supply are highly customizable.[1] Telemetry data are divided into three categories: *Cumulative, Delta,* and *Gauge*. The Cumulative category consists of integral values over time, such as total CPU time used; the Delta category represents a change over time, such as fluctuating bandwidth usage, while the Gauge category represents other discrete values recorded as a function of time.

For this work we have considered the following subset of the OpenStack meters provided by Ceilometer:

- CPU Utilization (gauge %)

- Disk Read Activity Bytes/Requests (gauge B/s, requests/s)
- Disk Write Activity Bytes/Requests (gauge B/s, requests/s)
- Network Incoming Bytes/Packets (gauge B/s, packets/s)
- Network Outgoing Bytes/Packets (gauge B/s, packets/s)
- Memory Usage (gauge, MB)

A total of ten meters were selected that we initially hypothesized would provide an accurate description of the status of our cloud in terms of applications being executed. We established a data pipeline between Ceilometer and Rabbit MQ, an open source messaging application. When the engineering applications are executed on any of the VMs provisioned in the cloud environment, the data pipeline is concurrently activated on a monitoring VM that outputs the telemetry data to a comma-separated value (CSV) file. In our study, we use a sampling rate of one sample every five seconds—higher than what is currently used by cloud metering systems for billing purposes, which is usually in the order of one sample every tens of minutes—to establish a baseline and to obtain the most accurate status of the cloud while the applications are running.

Smaller sampling rates will act as a smoothing filter on the time histories; this approach should be investigated for production clouds. It should be noted that the approach described in this paper can be generalized to cloud management systems other than OpenStack and telemetry data collectors such as Ceilometer. For example, the widely used monitoring system for HPC systems and Grids, Ganglia.[6] Due to the nature of the telemetry data collected, our approach can maintain the privacy-preserving features necessary in a production cloud environment. A more detailed description of privacy-preserving models for cloud computing can be found in Saxena and Pushkar.[7]

### III. Description of Telemetry Data

This section presents the telemetry data Ceilometer obtained running CADO-NFS, MCNP6, and OpenFOAM on the infrastructure we described. We will discuss the telemetry data as well as the characteristics of each computer code being tested.

#### A. MCNP6 Radiation Shielding Problems Telemetry Data

MCNP6 is a neutral- and charged-particle radiation transport code developed and maintained by Los Alamos National Laboratories (LANL).[3] The code was compiled with the MPI libraries to leverage the performance benefits provided by our HPC environment. In this study, we consider a sample problem commonly used in the nuclear engineering practice. This problem is part of the validation suite provided with MCNP6, and it represents a particular class of radiation transport problem defined as "fixed source" or "shielding problem," in which the input is a known radiation source, e.g., gamma or neutrons, and the application works to solve the radiation transport problem and compute the radiation dose within the computational domain. Another class of problems solved by MCNP6 is categorized as a "criticality eigenvalue" problem, in which the goal is to identify the source itself and ascertain the critical state of the system with regard to fission reactions. This second category of problem is not considered in this study, but is part of planned future work on the topic. In this study we have considered a shielding problem, referred to as BE08, consisting of solving a radiation transport problem involving a beryllium sphere with an outer radius of 12.58 cm.

BE08 was solved using MCNP6 running on both the *digi-a* and *digi-b* computing nodes (total of 16 processors). The telemetry data obtained on both nodes is shown in Figs. 1 through 4.

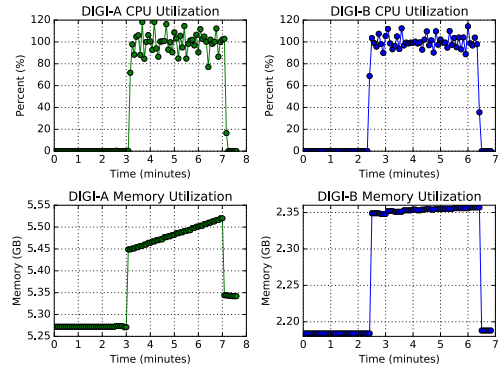

Fig. 1.  MCNP6 Telemetry Data – CPU and Memory Utilization

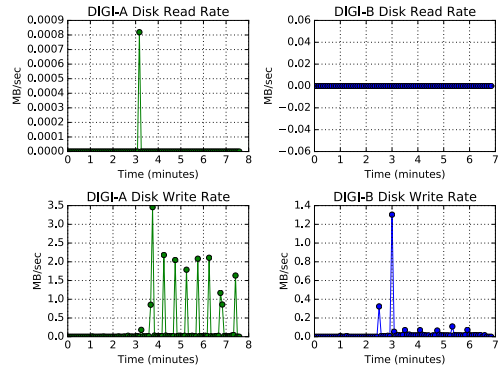

Fig. 2.  MCNP6 Telemetry Data – Disk I/O Activity (MB/sec)

Fig. 1 plots the CPU and memory utilization telemetry data. The load is equally distributed across both nodes with CPU utilization approaching 100%. The memory utilization exhibits a linear trend on *digi-a,* as the MCNP stores the particles histories (tallies) in memory. The second node, *digi-b*, displays a similar behavior but on a smaller scale; this is because *digi-a* is the master node and is used to store the statistical and particle history information. As Fig. 2 shows, disk read activity is almost zero on each node, with the exception of an initial spike when the code is started and the input parameters are read in memory. This is typical of an in-core type of application, which executes the entire computational process on the node's RAM. This feature makes the computer code very efficient because disk access (which is much slower than RAM) is limited. The disk write bytes and requests display a modest amount of activity as MCNP6 is saving the particle history, including statistical information at predetermined steps. As seen in Fig. 4, network activity is significant on both nodes due to the exchange of statistical information.

incompressible code that solves the coupled Maxwell-Navier-Stokes equations for an incompressible fluid (the fluid electrical conductivity is presumed constant). MHD is a field that has many applications in metallurgy, microfluidic pumping, fusion and fission reactor power generation, and cosmic studies.

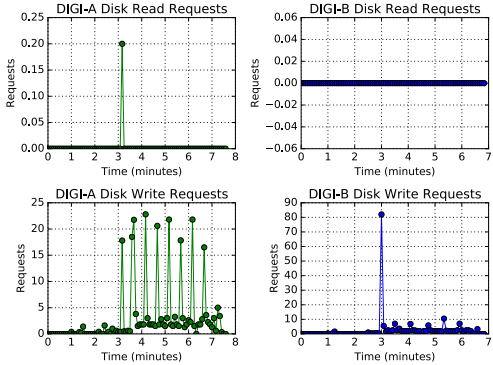

Fig. 3. MCNP6 Telemetry Data – Disk I/O Activity (Requests/sec)

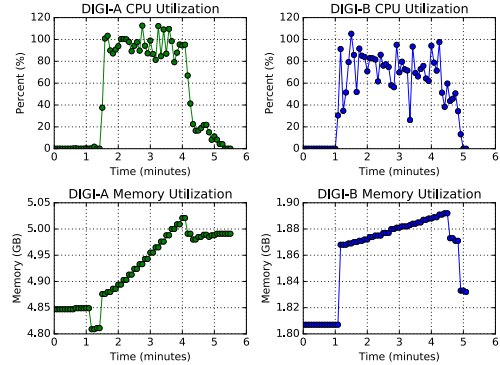

Fig. 5. OpenFOAM Telemetry Data – CPU and Memory Utilization

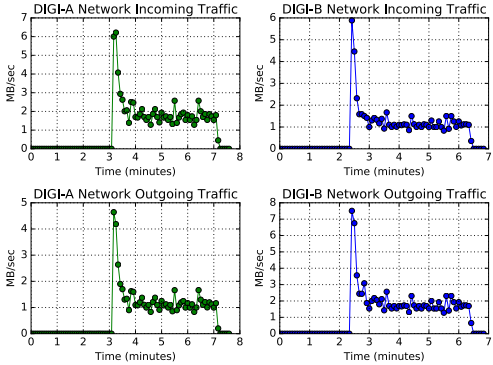

Fig. 4. MCNP6 Telemetry Data – Network Activity

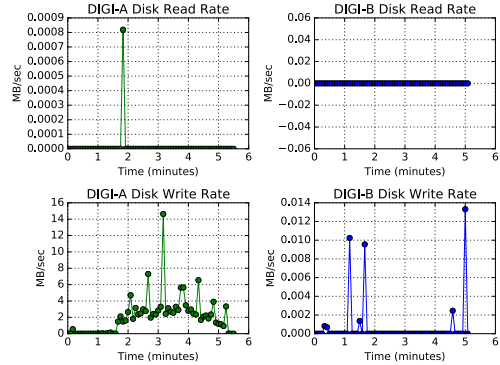

Fig. 6. OpenFOAM Telemetry Data – Disk I/O Activity (MB/sec)

## B. *OpenFOAM Magnetohydrodynamics (MHD) Telemetry Data*

OpenFOAM is a computational platform developed and maintained by OpenCFD Ltd., a subsidiary of the ESI group.[4] The software was originally developed for the numerical solution of computational fluid dynamics (CFD) problems. The tensorial notation paired with a highly flexible finite volume discretization of the spatial domain has transformed OpenFOAM into a tool capable of solving multiphysics problems, including heat transfer, electromagnetic fields, structural behavior, multiphase fluid dynamics, and other similar problems. Basically, any physical problem that can be described by a set of partial differential equations can be addressed with this tool.

OpenFOAM is capable of scaling to several hundreds of processors while maintaining adequate performance. We included it in our study because it is widely used for solving complex engineering problems such as aerodynamics modeling, combustion processes, complex multiphase modeling, and electromagnetic field simulation. Our work acquired telemetry data to solve a magnetohydrodynamics (MHD) problem using OpenFOAM *mhdFoam*, a currently

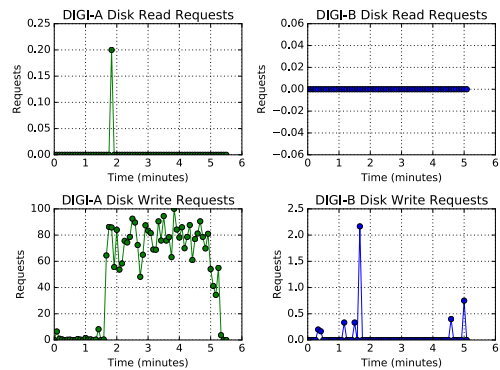

Fig. 7. OpenFOAM Telemetry Data – Disk I/O Activity (Requests/sec)

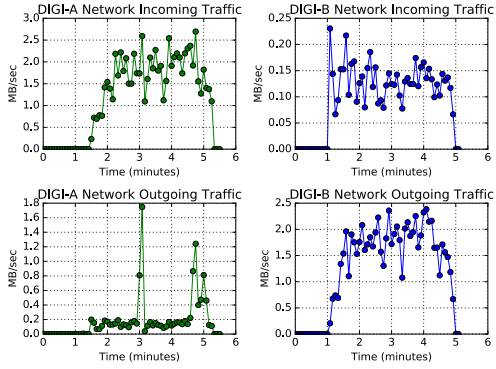

Fig. 8. OpenFOAM Telemetry Data – Network Activity

As Fig. 5 illustrates, the CPU and memory utilization metrics from OpenFOAM are somewhat similar to what was obtained using MCNP6. The spatial domain was partitioned on sixteen processors, using both the *digi-a* and *digi-b* nodes. The processor load balancing for this problem was not optimized using OpenFOAM's automatic load-balancing algorithm.

Figs. 6 and 7 demonstrate that OpenFOAM is an in-core application, since the disk read rate/requests are essentially zero, while the write rate/requests mainly occur on *digi-a*, as is typical of in-core applications. Fig. 8 shows significant network activity with peaks nearing 2.5 MB/sec between the two nodes. With OpenFOAM as with MCNP6, *digi-a* serves as the master node, collecting the solution information and storing it to disk. This activity underscores another distinguishing feature of applications developed for distributed memory architectures is that network communication is required at each iteration of the solution process to communicate the partial solution among the computing nodes.

## C. CADO-NFS Telemetry Data

In an effort to increase the variety of engineering applications considered in this study, we also analyzed CADO-NFS.[5] This computer code implements the Number Field Sieving (NFS) algorithm used for prime number factorization.

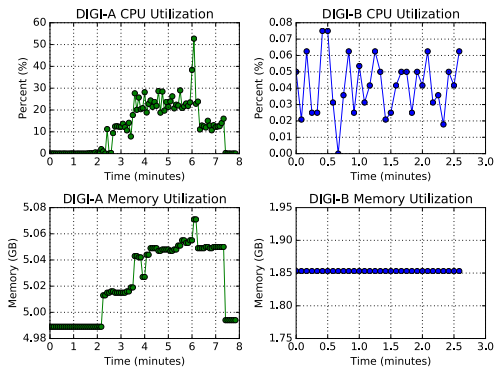

Fig. 9. CADO-NFS Telemetry Data – CPU and Memory Utilization

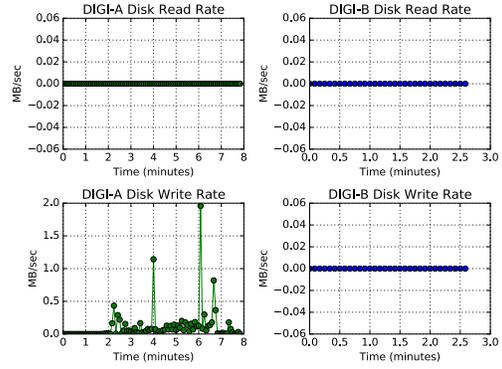

Fig. 10. CADO-NFS Telemetry Data – Disk I/O Activity (MB/sec)

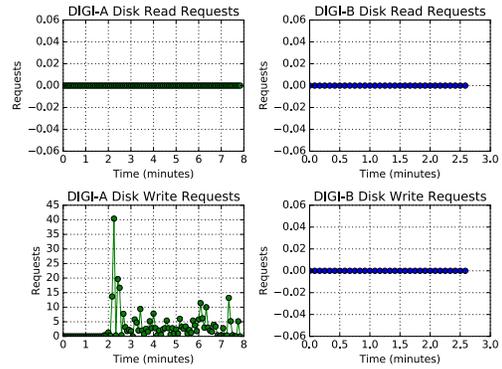

Fig. 11. CADO-NFS Telemetry Data – Disk I/O Activity (Requests/sec)

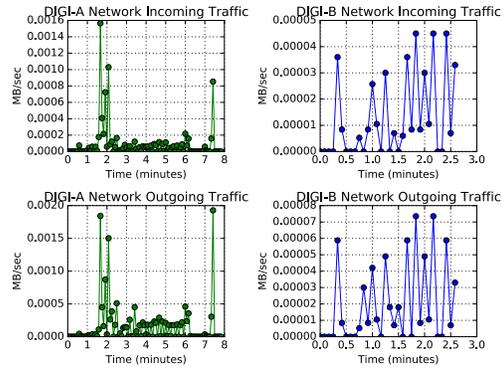

Fig. 12. CADO-NFS Telemetry Data – Network Activity

Figs. 9 through 12 clearly indicate that CADO-NFS is running on the *digi-a* node only. While the telemetry plots for the *digi-b* node show only minimal activity, the CPU and memory utilization metrics indicate complex activity from the different steps required to factor a prime number. The spike in CPU utilization around the sixth minute of execution corresponds to the one of the most computationally intensive step of the process: solving a linear system of equations.

## IV. IDENTIFICATION OF ACTIVE APPLICATIONS WITH MACHINE LEARNING ALGORITHMS

This section describes the approach we took in using machine learning algorithms to identify applications. We used the telemetry data and features described in Section II as inputs for several types of of classifiers. Training and validation data sets were independently obtained by multiple executions of the applications. The telemetry data obtained from both *digi-a* and *digi-b* were used to train and validate the classifiers. We observed a performance improvement from using data from both nodes. Also, in a production environment we expect that the monitoring system will feed data to the classifiers from every VM instance. The machine learning algorithms were developed using the *scikit-learn*[2] package available for the Python language.

The machine learning algorithms developed and tested in this work include the following classifiers:

- Decision Trees (DT)
- *k*-Nearest Neighbor (kNN)
- Support Vector Machines (SVM)
- Multi Layer Perceptron Neural Networks (MLP-NN)

Parameter settings for these models were chosen by using a grid-search algorithm to optimize precision and recall metrics. The precision metric indicates the *percentage* of relevant or true positive hits of the application running versus the total of true and false positives. The recall metric indicates the *fraction* of true positives versus the total of true positives and false negatives.

The search space for each classifier's parameters is listed in Table I. The final values are listed in Table II. Minimal optimization was performed on the MLP-NN; this is part of future work we plan to perform on this topic.

TABLE I. GRID-SEARCH PARAMETERS SPACE

| Classifier | Search Space | |
| --- | --- | --- |
| | Parameter | Range |
| **DT** | Tree depth | [1…20] |
| **kNN** | Number of Neighbors | [1…20] |
| | Weights | Uniform / Distance |
| | Algorithm | ball-tree / kd-tree |
| **SVM** | Kernel | Radial Basis Functions / Linear |
| | gamma | [1e-3, 1e-4] |
| | C | [1, 10, 100, 1000] |

Two independent data sets were developed, a training data set and a validation data set. Grid-search parameter optimization was performed using the training set. Our evaluation metrics, such as confusion matrices and scoring statistics, were evaluated based on the validation data set. For each data set, each program was run for a specific scenario, MCNP6 problem BE08, the OpenFOAM MHD application, and CADO-NFS.

The grid-search optimization performed on the data set in this scenario yields the optimal parameters listed in Table II. The MLP-NN consists of three hidden layers with 16, 8, and 4 neurons for each layer. A sigmoid threshold function was specified for the first two hidden layers and a linear function was specified for the third layer. A final softmax layer was used to generate the classification output. The learning rate of 0.01 was specified and the Stochastic Gradient Descent learning algorithm was used. Normalized confusion matrices and scoring statistics for each classifier are listed in Tables III through VI.

TABLE II. OPTIMAL PARAMETERS SPACE

| Classifier | Search Space | |
| --- | --- | --- |
| | Parameter | Range |
| DT | Tree depth | 10 |
| kNN | Number of Neighbors | 5 |
| | Weights | Uniform |
| | Algorithm | Ball-tree |
| SVM | Kernel | Linear |
| | gamma | 1e-3 |
| | C | 1000 |

TABLE III. DECISION TREE (DT) NORMALIZED CONFUSION MATRIX

| | | Predicted Label | | |
| --- | --- | --- | --- | --- |
| | | CADO-NFS | MCNP6 | OpenFOAM |
| **True Label** | CADO-NFS | **0.71** | 0.0 | 0.29 |
| | MCNP6 | 0.0 | **0.99** | 0.01 |
| | OpenFOAM | 0.0 | 0.02 | **0.98** |

DT SCORING STATISTICS

| | Precision | Recall | F1-score |
| --- | --- | --- | --- |
| **CADO-NFS** | 1.0 | 0.71 | 0.83 |
| **MCNP6** | 0.96 | 0.97 | 0.97 |
| **OpenFOAM** | 0.80 | 0.91 | 0.85 |

TABLE IV. K-NEAREST NEIGHBOR (kNN) NORMALIZED CONFUSION MATRIX

| | | Predicted Label | | |
| --- | --- | --- | --- | --- |
| | | CADO-NFS | MCNP6 | OpenFOAM |
| **True Label** | CADO-NFS | **1.0** | 0.0 | 0.0 |
| | MCNP6 | 0.0 | **0.98** | 0.02 |
| | OpenFOAM | 0.19 | 0.04 | **0.77** |

KNN SCORING STATISTICS

| | Precision | Recall | F1-score |
| --- | --- | --- | --- |
| **CADO-NFS** | 0.75 | 1.0 | 0.86 |
| **MCNP6** | 0.98 | 0.98 | 0.98 |
| **OpenFOAM** | 0.94 | 0.77 | 0.85 |

TABLE V. SUPPORT VECTOR MACHINES (SVM) NORMALIZED CONFUSION MATRIX

|  |  | Predicted Label | | |
|---|---|---|---|---|
|  |  | CADO-NFS | MCNP6 | OpenFOAM |
| True Label | CADO-NFS | **1.0** | 0.0 | 0.0 |
|  | MCNP6 | 0.0 | **0.97** | 0.03 |
|  | OpenFOAM | 0.18 | 0.32 | **0.5** |

SVM SCORING STATISTICS

|  | Precision | Recall | F1-score |
|---|---|---|---|
| CADO-NFS | 0.76 | 1.0 | 0.86 |
| MCNP6 | 0.87 | 0.97 | 0.92 |
| OpenFOAM | 0.90 | 0.50 | 0.64 |

TABLE VI. MULTI LAYER PERCEPTRON NEURAL NETWORK (MLP-NN) NORMALIZED CONFUSION MATRIX

|  |  | Predicted Label | | |
|---|---|---|---|---|
|  |  | CADO-NFS | MCNP6 | OpenFOAM |
| True Label | CADO-NFS | **1.0** | 0.0 | 0.0 |
|  | MCNP6 | 0.0 | **0.93** | 0.07 |
|  | OpenFOAM | 0.18 | 0.05 | **0.77** |

MLP-NN SCORING STATISTICS

|  | Precision | Recall | F1-score |
|---|---|---|---|
| CADO-NFS | 0.76 | 1.0 | 0.86 |
| MCNP6 | 0.98 | 0.93 | 0.95 |
| OpenFOAM | 0.83 | 0.77 | 0.80 |

The DT algorithm performed best in this scenario; kNN, SVM, and MLP-NN tended to confuse CADO-NFS with OpenFOAM. kNN and MLP-NN exhibited nearly matched accuracy. As already mentioned, limited optimization was performed on the MLP-NN algorithm. The normalized confusion matrices for each classifier are graphically presented in Fig. 13 along with the relative score for each algorithm.

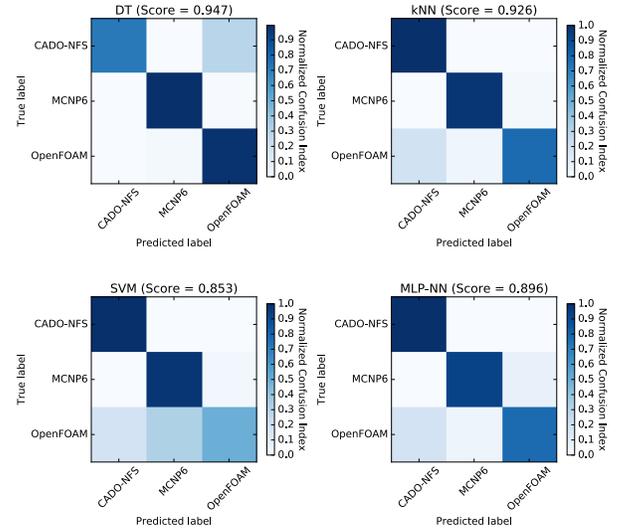

Fig. 13. Plot of the Normalized Confusion Matrix for Different Classifiers

### A. Improving Model Predictions Using Principal Component Analysis (PCA) for Case 1

We decomposed the training and validation data sets using Principal Component Analysis (PCA)[8] to evaluate the possibility of using the orthogonal components of the transformed space to improve the classification performance. In mathematics and statistics, this type of approach can be categorized as *regularization*, and it is a process whereby additional data is introduced to solve ill-conditioned problems or to lessen overfitting. In addition, the PCA decomposition is used to identify potential clustering of the data sets in the transformed space that can be used to provide additional features to train the ML classifiers.

We computed the first three components of the PCA decomposition for the training data set. The explained variance obtained was 0.58, 0.28, and 0.1, for the first, second, and third principal components, respectively. We found that increasing the number of principal components did not yield significant improvements in the explained variance; therefore, we truncated the PCA decomposition after the third component. Plots of the principal components are shown in Figs. 14 through 16. These plots show interesting patterns emerging from the PCA decomposition of the training data set. In particular, clusters can be distinguished for both MCNP6 and CADO-NFS.

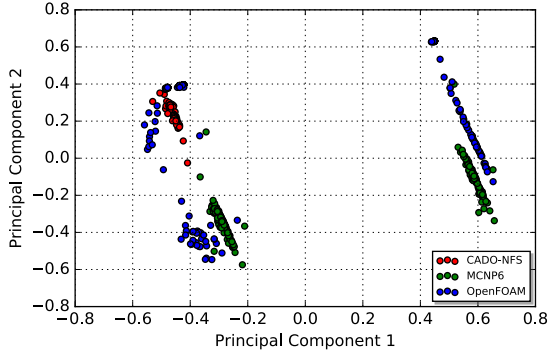

Fig. 14. Plot of PCA Components 1 and 2 for the Training Data Set

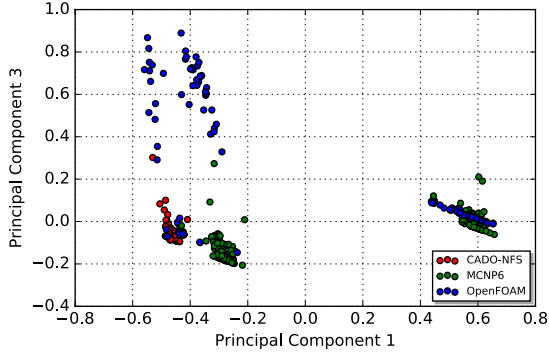

Fig. 15. Plot of PCA Components 1 and 3 for the Training Data Set

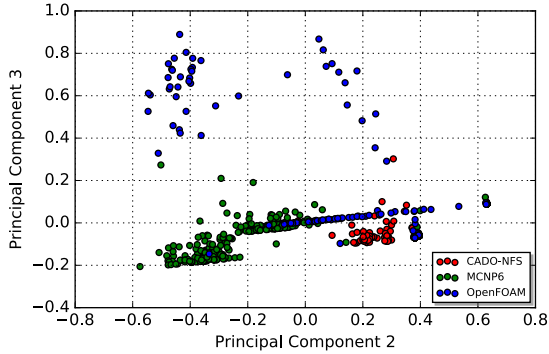

Fig. 16. Plot of PCA Components 2 and 3 for the Training Data Set

The first two components of the PCA decomposition were used to augment the training and validation data sets. We observed both slight performance improvements and slight degradations across the classifiers. Tables VII through X show the scoring statistics using raw data, with the PCA-augmented data sets distinguished by square brackets […]. The PCA-augmented features yielded a maximum performance improvement of around 4%; however, performance was lowered by 5% in some cases. We believe that the limited improvements are due to the already high performance achieved with the original data sets. We expect to obtain more conclusive results on the effect of the PCA decomposition with further testing.

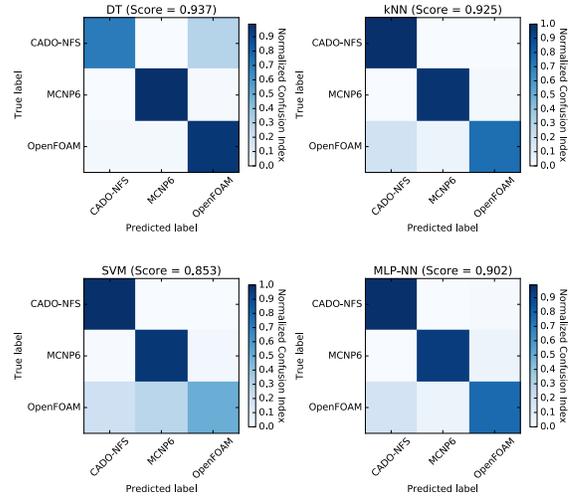

Fig. 17. Case 1 - Plot of the Normalized Confusion Matrix for Different Classifiers Trained and Validated with the PCA-Augmented Data Set

TABLE VII.  DT SCORING STATISTICS

|  | Precision | Recall | F1-score |
|---|---|---|---|
| **CADO-NFS** | 1.0 [0.95] | 0.71 [0.70] | 0.83 [0.81] |
| **MCNP6** | 0.96 [0.98] | 0.97 [0.99] | 0.97 [0.99] |
| **OpenFOAM** | 0.80 [0.84] | 0.91 [0.95] | 0.85 [0.89] |

TABLE VIII.  kNN SCORING STATISTICS

|  | Precision | Recall | F1-score |
|---|---|---|---|
| **CADO-NFS** | 0.75 [0.75] | 1.0 [1.0] | 0.86 [0.86] |
| **MCNP6** | 0.98 [0.98] | 0.98 [0.98] | 0.98 [0.98] |
| **OpenFOAM** | 0.94 [0.95] | 0.77 [0.76] | 0.85 [0.84] |

TABLE IX.  SVM SCORING STATISTICS

|  | Precision | Recall | F1-score |
|---|---|---|---|
| **CADO-NFS** | 0.76 [0.72] | 1.0 [1.0] | 0.86 [0.84] |
| **MCNP6** | 0.87 [0.88] | 0.97 [0.97] | 0.92 [0.92] |
| **OpenFOAM** | 0.90 [0.90] | 0.50 [0.50] | 0.64 [0.65] |

TABLE X.  MLP-NN SCORING STATISTICS

|  | Precision | Recall | F1-score |
|---|---|---|---|
| **CADO-NFS** | 0.76 [0.75] | 1.0 [0.99] | 0.86 [0.86] |
| **MCNP6** | 0.98 [0.98] | 0.93 [0.94] | 0.95 [0.96] |
| **OpenFOAM** | 0.83 [0.85] | 0.77 [0.77] | 0.80 [0.81] |

## V. Conclusions and Future Work

We developed a methodology for the identification of engineering applications running in a high-performance cloud environment. The Ceilometer metering tool obtains data by passively logging cloud activity (such as CPU utilization, disk activity, network activity, and many more activity types) while maintaining user privacy. The cloud environment in these experiments was configured as an HPCaaS infrastructure, similar to what can be achieved with a classic Beowulf cluster. Two computing nodes were configured to share the same disk parition via NFS and communicate via a high-speed network interconnect. We examined three applications of interest to nuclear engineering practice and other technical areas, namely MCNP6, OpenFOAM, and CADO-NFS.

Our results indicate that machine learning classifiers can achieve a detection accuracy greater than 90% with the data sets analyzed. Decision tree classifiers and Multi-Layer Perceptron Neural Networks yield the best classification performance for our problem. Prior work has supported the feasibility of program recognition on individual virtual machines, and this work expands that capability to multi-node programs. Our future work will include the development of a database of telemetry data that spans a larger set of input parameters and test scenarios for the applications of interest. In addition, we plan to perform further optimization of the machine learning classifiers, with particular focus on deep neural networks.

In this paper, we have described a methodology that supports detection of a large number of distinct scientific applications in real-time in an HPCaaS type of cloud. Finally, we plan to build on these results to develop a family of "detection metrics." We expect that these metrics will provide quantitative information on various classes of scientific applications running on the cloud that may require further investigation.